\documentclass{moriond}

\usepackage{float}
\def\be{\begin{equation}}
\def\ee{\end{equation}}
\def\bea{\begin{eqnarray}}
\def\eea{\end{eqnarray}}

\def\MM{M_{pl}}
\newcommand{\utsr}{^{\mu\nu}}               %up tensor
\newcommand{\Tsr}{_{IJ}}                    %tensor in phase space
\newcommand{\uTsr}{^{IJ}}                   %up tensor in phase space
\newcommand{\sR}{{\cal R}}
\usepackage{stmaryrd}
\usepackage{amsmath}
\usepackage{multicol}
\newcommand{\Photo}{\includegraphics[height=0.01mm]{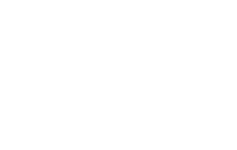}}

\begin{document}
\vspace*{4cm}
\title{The Separate Universe approach for multifield inflation models}

\author{ Hugo Holland }

\address{Institut d'Astrophysique Spatiale, rue Jean Teillac \\
91405 ORSAY CEDEX, France}

\maketitle\abstracts{Primordial black holes could constitute part or all of dark matter but they require large inhomogeneities to form in the early universe. These inhomogeneities can strongly backreact on the large scale dynamics of the universe. Stochastic inflation provides a way of studying this backreaction and getting an estimation of the abundance of primordial black holes. Because stochastic inflation focuses on large scale dynamics, it rests on the separate universe approach. However, the validity of this approach has only been checked in single field models, but not in multifield models in which we expect strong boosts in the power spectrum, leading to the formation of primordial black holes. We will check the validity of a separate universe approach in multifield models by matching it with a complete cosmological perturbation theory approach at large scales. In particular, we wish to compare these two paradigms and their differences in the adiabatic and entropic directions of the phase space. This will give us a range of validity and conditions one needs to verify in order to apply the separate universe approach and stochastic inflation in multifield models.}

\section{Stochastic inflation \& Separate Universe}
    The search for dark matter candidates has given rise to many suitable candidates. One of these are primordial black holes (PBH). These black holes are different than astrophysical black holes as they did not form after the collapse of an object such as a star, but formed from the collapse of overly dense regions in the primordial universe. We now have many constraints on the proportion of dark matter that PBH could account for depending on their masses \cite{Carr_2021}. However some mass regions are still completely open and there is still hope to detect a gravitational wave background coming from PBH with LISA \cite{Bartolo_2019}. There are several ways of ensuring the creation of PBH in inflationary models, but all of them boil down to the same issue : creating a massive boost in the power spectrum. These boost can appear in both single field models of inflation and multifield models \cite{Cole:2023wyx} \cite{Geller_2022}. In any model, these boosts will produce very large fluctuations and will impact the dynamics of large scale fluctuations, which in turn will impact the abundance of collapsed structures such as PBHs or dark matter haloes \cite{Ezquiaga:2022qpw}. The way to study these dynamics is the Stochastic Inflation formalism \cite{vennin2022stochastic}.
    \subsection{Stochastic inflation}
    
\begin{figure}[H]
        \centering 
           \includegraphics[scale = 0.65]{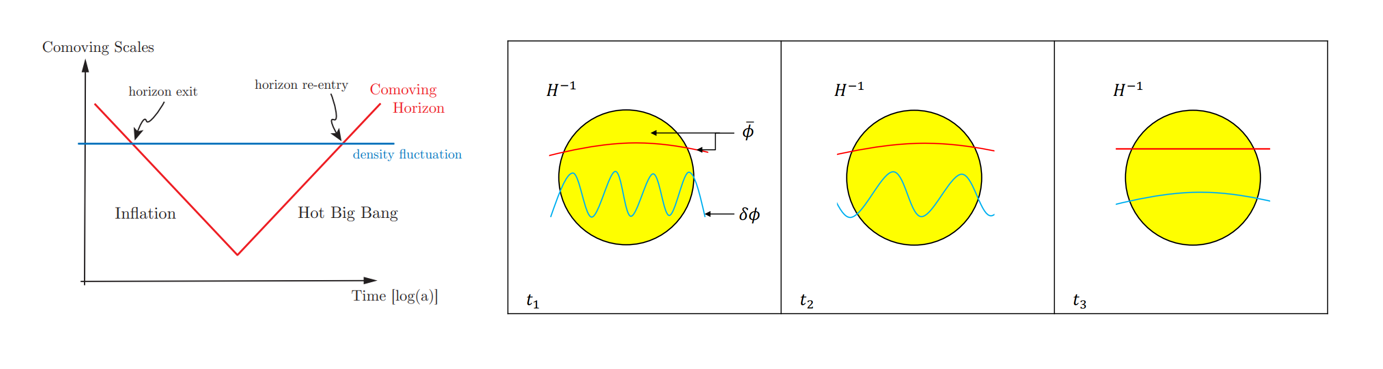}
           \caption{Solution to the horizon problem via inflation. \href{https://arxiv.org/abs/0907.5424}{\underline{\footnotesize	{Bauman TASI lectures}}}(left) And Quantum to classical transition of fluctuations \href{https://www2.yukawa.kyoto-u.ac.jp/~gc2024/slides/240129-2_Artigas.pdf}{\underline{\footnotesize	{Artigas Grav\&Cosmo 2024}}} (right) }
           \label{baumann}
        \end{figure}

        We know that during inflation the Hubble radius $(aH)^{-1}$ decreases. This means that for a given field $\phi$, each mode $\phi_k$ is either sub horizon or super horizon. During inflation there will be a continuous flow of sub horizon modes crossing into the super horizon regime as seen in \ref{baumann}(left).
        \noindent This continuous flow of subhorizon modes crossing the horizon, is treated as a white noise in the dynamics of the super horizon modes, giving a Langevin equation. We will denote the super horizon modes as $\phi_{IR}$.
        \begin{equation}
            \dot\phi_{IR} = -\frac{\partial V}{\partial \phi} + \xi_\phi
        \end{equation}
        Where $\xi_\phi$ is the noise that depends on the sub horizon scales.
        In this equation we can immediately notice that there are no gradients, so no space dependence. That is because we have assumed that each Hubble patch is homogeneous and isotropic and that they all evolve independently from one another. This is the separate universe approach \cite{Artigas_2022} \cite{Pattison_2019}.

    \subsection{The separate universe approach}

        We will now have two different approaches in order to treat the different regimes. Having lost the inhomogeneity for the super horizon modes, we can study them at a non linear level. On the contrary the sub horizon modes evolve in a inhomogeneous background so we will restrict ourselves to a linear evolution. The evolution of a given mode compared to the background is given in \ref{baumann} (right). In stochastic inflation we allow a abrupt transition from one regime to the other as the mode crosses the horizon. This abrupt transition is exctly what the separate universe approach aims to justify.

    \section{Separate universe \& multifield inflation}

        In this section we will present a very generic model of multifield inflation, often referred to as "non linear sigma models" and how the separate universe approach affects such models.

        \subsection{Multifield inflation}
            We start from a very generic Lagrangian for multifield models of inflation
            with $I{,}J \in \llbracket 1,n\rrbracket$ is
            \begin{equation}
                S = \int d^4x \sqrt{-g} \big [\frac 1 2 \MM^2 \sR - \frac 1 2 g\utsr G\Tsr \partial_\mu \phi^I \partial_\nu \phi^J - V(\phi^I) \big]
            \end{equation}
            Where $\phi^I$ are our fields and $G\uTsr$ is a coupling metric that depends on all of the fields. It is interesting to note that at the classical level these models are equivalent to non minimally coupled models in which $G\uTsr = \delta\uTsr$ and $\sR$ is replaced by an $f(\sR)$ function. This model doesn't only encompass non minimal couplings to gravity, it can also represent non linear couplings between the fields or derivative couplings like those that appear in EFT of inflation. Many models of multifield high energy physics can be generalised with this metric coupling.
            We then make use of the ADM formalism and perturb all of our objects in two ways. Firstly we do a complete cosmological perturbation theory development, then we do the same thing but with the separate universe assumption.

        \subsection{Validity conditions of the Separate Universe Approach}
            Once everything has been perturbed at the second order we can compare the equations of motions found in the full CPT approach at large scales to those found in the SU approach and check when they are consistent. This is equivalent to comparing the two Hamiltonians that we get :
            \begin{align}
                H_{CPT} &= \int d^3 \vec x \big [\delta N \mathcal S^{(1)} + \delta N^i \mathcal D_i + N\mathcal H^{(2)}\  \big ] \\
                H_{SU} &= \int d^3 \vec x \big [ \bar{\delta N} \bar{\mathcal S}^{(1)}  + \bar N\bar{\mathcal H }^{(2)} \big ] 
            \end{align}
            Where a bar means that we are working in the separate universe formalism, $N$ and $N_i$ are the lagrange multipliers that come from the ADM formalism, $S^{(1)}$ and $H^{(2)}$ are the scalar constraint at first and at second order in perturbation, and $D_i$ is the diffemorphism contraint expressed at first order in perturbation. We finally compare $S^{(1)}$ with $\bar S^{(1)}$ and $H^{(2)}$ with $\bar H^{(2)}$ and we get three inequalities that need to be verified in order to justify the separate universe approach. The results we present here are twofold. On the left column we give the inequalities as we found them in a fully covariant expression of the hamiltonians expressed in the original field space. In the column on the right we give the same inequalities but expressed in the adiabatic and entropic directions of our field space \cite{Gordon_2000}.
            \begin{align}
                 \MM^2\big( \frac{k}{aH} \big )^2 &\ll \frac{1}{H^2} \big | \frac{\pi_I\pi^I}{a^6} - V(\phi^I) \big | \quad=\quad \frac{1}{H^2}\big | N^2\dot\sigma^2- V(\phi^I) \big |\\
                \MM^2\big( \frac{k}{aH} \big )^2 &\ll \frac{1}{H^2} \big | \frac{\pi_I\pi^I}{a^6} + \frac{V(\phi^I)}{2} \big | \quad=\quad \frac{1}{H^2} \big | N^2\dot\sigma^2 + \frac{V(\phi^I)}{2} \big |\\
                \big |\big ( \frac{k}{aH} \big )^2 G\Tsr \big | &\ll  \frac{1}{H^2} \big | \mathcal M\Tsr \big | \quad\quad\quad\quad\Leftrightarrow \quad\quad \big ( \frac{k}{aH} \big )^2 \ll  \frac{1}{H^2} \big | \mathcal M_{cc} \big |
            \end{align}
            The first two inequalities tell us that we have to be at big enough scales for the separate universe approach to work, as expected because we were aiming for scales at which the hubble patches can be seen as homogeneous and isotropic. This isn't quite obvious to see but the right hand side is essentially Friedmann's equations which once divided by $\MM^2$ gives one. The third inequality is slightly more subtle as it tell us that the effective mass of our fields have to be large enough. The adiabatic and entropic versions of the third inequality show that the condition on the mass has to be valid for all the directions. However, we expect the adiabatic direction to be the hardest to constrain as it is the flatest and lightest directions. We are still studying these directions in this approach as for more than three fields it is very easy to constrain the first derivatives of the potential with respect to the entropic ones \cite{Pinol:2020kvw}.
    \section{Gauge fixing and Mukhanov Sasaki variables}
    One point that we have not yet discussed here is the issue of gauge fixing. It has been shown that there are issues in some gauges when going from the Separate universe framework to the full CPT framework as some gauge degrees of freedom don't exist in the former so cannot be easily fixed. This is fairly simple to see when computing the equations of motions for the Mukhanov-Sasaki gauge invariant variables in the separate universe approach, in particular for the adiabatic direction :
    \begin{align}
        \ddot {\bar Q}_\sigma + 3H\dot{\bar{Q}}_\sigma + \big[\mathcal M_{\sigma\sigma} - \omega_1^2 - \frac{1}{\MM^2a^3}&\partial_t(\frac{a^3\dot\sigma^2}{H})\big] \bar{Q}_\sigma=\mathcal F(\bar{Q}_1, \dot{ \bar{Q}}_1) + \frac{3}{2}{\MM^2 a^2}\dot\sigma\bar{\mathcal D_i}.
    \end{align}
    Where $\omega_1$ is the covariant rate turn, $\mathcal M_{\sigma\sigma}$ is the effective mass of the adiabatic Mukhanov-Sasaki variable and $\mathcal F(Q_1, \dot Q_1)$ gives the coupling between the adiabatic direction and the first entropic direction. In the full CPT computation we get an extra $k^2 \bar{Q}_\sigma$ term in the left hand side and no $\bar D_i$ term in the right hand side, which corresponds to an additional source term coming from a diffeomorphism constraint even though such a constraint doesn't naturally appear in the hamiltonian formalism in the SU approach \cite{Artigas_2022}.

    \section{Conclusion}

        The separate universe approach works as long as we are being careful with it. We need to work at large enough scales or guaranty that any potential $k^2$ terms are suppressed in front of the effective mass of our fields. The gauge fixing issues that appear here are the same than in single field models, as long as we start from the CPT framework we won't have any problems, but one needs to be careful when fixing a gauge in the SU framework and trying to get back to a more general framework as new degrees of freedom might still be unfixed.
        
    \section*{Acknowledgments}
    
        I would like to thank Julien Grain and Lucas Pinol for the many enlightening conversations I regularly have with them.

\section*{References}

\bibliography{ref}
\end{document}